\documentclass[12pt]{article}

\usepackage{graphicx}
\begin{document}

\begin{center}
{\bf Rational nonlinear electrodynamics of AdS black holes and extended phase space thermodynamics } \\
\vspace{5mm} S. I. Kruglov
\footnote{E-mail: serguei.krouglov@utoronto.ca}
\underline{}
\vspace{3mm}

\textit{Department of Physics, University of Toronto, \\60 St. Georges St.,
Toronto, ON M5S 1A7, Canada\\
Department of Chemical and Physical Sciences, University of Toronto,\\
3359 Mississauga Road North, Mississauga, Ontario L5L 1C6, Canada} \\
\vspace{5mm}
\end{center}
\begin{abstract}
The critical behaviour of magnetically charged by rational nonlinear electrodynamics (RNED) AdS black holes in an extended phase space is investigated. The cosmological constant is considered as a thermodynamic pressure and the black hole mass is
 identified with the chemical enthalpy. The analogy with the Van der Walls liquid–gas system is found and the critical exponents coincide with those of the Van der Waals system. The thermodynamics of RNED-AdS black holes and phase transitions are studied.
A new thermodynamic quantities conjugated to the nonlinear parameter of RNED and a magnetic charge are defined. It is demonstrated the  consistency of the first law of black hole thermodynamics and the Smarr formula.
\end{abstract}

\section{Introduction}

Nowadays black holes are treated as a thermodynamics system \cite{Bardeen}, \cite{Jacobson}, \cite{Padmanabhan}  with the area of a black hole being the entropy and its surface gravity is identified with the temperature \cite{Bekenstein}, \cite{Hawking}. This can give the progress in understanding of the link between gravity and quantum physics. The consideration of anti-de Sitter (AdS) space-time with a negative cosmological constant led to the phase  behaviour of black holes \cite{Page}. The important step was made later to consider a holographic picture where black holes are a system which is dual to conformal field theories \cite{Maldacena}, \cite{Witten}, \cite{Witten1}. The studding the holography gave rise to a progress in solving problems in quantum chromodynamics \cite{Kovtun}, and condensed matter physics \cite{Kovtun1}, \cite{Hartnoll}. It was proposed recently that the cosmological constant plays a role of a pressure which is a conjugate to a volume in the black hole thermodynamics. This allowed to compare the phase transitions in black holes with those in liquid-gas thermodynamics \cite{Dolan}, \cite{Kubiznak}, \cite{Mann}, \cite{Teo}. Some aspects of Born--Infeld (BI) electrodynamics in AdS space-time with the negative cosmological constant were studies in \cite{Fernando}-\cite{Miskovic}. It was found an analogy to Van der Waals fluids in the black hole physics.
In this paper we study the thermodynamics in the framework of the rational nonlinear electrodynamics (RNED) \cite{Kr0}, \cite{Kr2} coupled to gravity with the negative cosmological constant. In the preceding studies \cite{Kr2} it was considered the thermodynamics in the asymptotically flat space-time in the framework of Einstein's theory of relativity without the cosmological constant. As a result, the pressure was not introduced, the P-V term was absent in black hole thermodynamics and there was not the analogy with Van der Vaals fluid. In addition, in \cite{Kr2} the local stability of black holes for the canonical ensemble (the charge is fixed) by calculating the specific heat was investigated. In the present paper we study the global black hole stability by analysing the Gibbs free energy within the grand canonical ensemble in the extended phase space. The model of nonlinear electrodynamics RED possesses an attractive properties such as the absence of singularity of charges at the origin and the finite of the electrostatic energy. Similar features take place in Born--Infeld electrodynamics. In addition, RNED coupled to gravity describes the inflation of the universe \cite{Kr} and provides the correct shadow of M87* black hole \cite{Kr1}. Electrically charged black holes within RNED coupled to gravity without the cosmological constant were studied in \cite{Kr3}. Here, we investigate magnetically charged black holes in the framework of gravity with the cosmological constant.

The structure of the paper is as follows. In Sec. 2 we obtain the RNED-AdS metric function with the asymptotic as $r\rightarrow\infty$ and $r\rightarrow 0$. When the Schwarzschild mass is zero the solution is non-singular with the de Sitter core as $r\rightarrow 0$. The extended thermodynamic phase space including a negative cosmological constant as a pressure with the conjugate volume and coupling $\beta$ of RNED was analyzed in Sec. 3. We found the thermodynamic magnetic potential and the thermodynamic conjugate to the coupling. The generalized
Smarr relation was obtained. The black hole thermodynamics was studied in Sec. 4. We obtained critical values of the specific volume, critical temperature and critical pressure. The Gibbs free energy is analysed in subsection 4.1. The black hole mass is considered as a chemical enthalpy. We depicted the pressure and the critical isotherms. In subsection 4.2 the critical exponents were established. Section 5 is a summary.

We use units with $c=1$, $\hbar=1$, $k_B=1$.

\section{RNED-AdS solution}

The action of Einstein–RNED theory in AdS space-time is given by
\begin{equation}
I=\int d^{4}x\sqrt{-g}\left(\frac{R-2\Lambda}{16\pi G_N}+\mathcal{L}(\mathcal{F}) \right),
\label{1}
\end{equation}
where the negative cosmological constant is $\Lambda=-3/l^2$ and $l$ being AdS radius. The RNED Lagrangian \cite{Kr0} in Gaussian units reads
\begin{equation}
{\cal L}(\mathcal{F}) =-\frac{{\cal F}}{4\pi(1+2\beta{\cal F})},
\label{2}
\end{equation}
with ${\cal F}=F^{\mu\nu}F_{\mu\nu}/4=(B^2-E^2)/2$, $F_{\mu\nu}=\partial_\mu A_\nu-\partial_\nu A_\mu$. The maximum electric field in the origin is $E(0)=1/\sqrt{\beta}$ \cite{Kr0}.
By varying action (1) with respect to $g_{\mu\nu}$ and $A_{\mu}$, one can obtain the equations of gravitational and electromagnetic fields as follows
\begin{equation}
R_{\mu\nu}-\frac{1}{2}g_{\mu \nu}R+\Lambda g_{\mu \nu} =8\pi G_N T_{\mu \nu},
\label{3}
 \end{equation}
\begin{equation}
\partial _{\mu }\left( \sqrt{-g}\mathcal{L}_{\mathcal{F}}F^{\mu \nu}\right)=0,
\label{4}
\end{equation}
where $R$  is the Ricci scalar and $R_{\mu \nu }$ is the Ricci tensor. The energy-momentum symmetric tensor of electromagnetic fields can be find by the variation of $\mathcal{L}\left(\mathcal{F}\right)$ with respect to the metric tensor and it is given by
\begin{equation}
 T_{\mu\nu }=F_{\mu\rho }F_{\nu }^{~\rho }\mathcal{L}_{\mathcal{F}}+g_{\mu \nu }\mathcal{L}\left( \mathcal{F}\right),
\label{5}
\end{equation}
and $\mathcal{L}_{\mathcal{F}}=\partial \mathcal{L}( \mathcal{F})/\partial \mathcal{F}$.

We consider the four-dimensional static spherical symmetric line element
\begin{equation}
ds^{2}=-f(r)dt^{2}+\frac{1}{f(r)}dr^{2}+r^{2}\left( d\theta
^{2}+\sin ^{2}\theta d\phi ^{2}\right).
\label{6}
\end{equation}
The spherical symmetry leads to the tensor $F_{\mu\nu}$ which involves only the radial electric field $F_{01}=-F_{10}$ and radial
magnetic field $F_{23}=-F_{32}=q_m\sin(\theta)$, where $q_m$ is the magnetic charge. The energy-momentum tensor becomes diagonal with $T_{0}^{~0}=T_{r}^{~r}$ and $T_{\theta}^{~\theta}=T_{\phi}^{~\phi}$. The metric function, within GR, with the spherical symmetry is given by
\begin{equation}
f(r)=1-\frac{2m(r)G_N}{r},
\label{7}
\end{equation}
where the mass function is \cite{Bronnikov}
\begin{equation}
m(r)=m_0+4\pi\int_{0}^{r}\rho (r)r^{2}dr,
\label{8}
\end{equation}
where $m_0$ is the integration constant corresponding to the Schwarzschild mass, $\rho$ is the energy density which also includes the term due to the cosmological constant. In the following we consider only magnetic  black holes because the electrically charged black holes (for models which have Maxwell's weak-field limit) leads to singularities \cite{Bronnikov}.

Now, we will study the static magnetic black holes. Taking into account that the
electric charge $q_e=0$, $\mathcal{F}=q_m^2/(2r^4)$ ($q_m$ is a magnetic
charge), we obtain from Eq. (5) the magnetic energy density plus the term corresponding to the negative cosmological constant
\begin{equation}
\rho=\frac{q_m^2}{8\pi(r^4+\beta q_m^2)}-\frac{3}{8\pi G_Nl^2}.
\label{9}
\end{equation}
With the help of Eqs. (8) and (9) one finds the mass function
\[
m(r)=m_0+\frac{q_m^{3/2}}{8\sqrt{2}\beta^{1/4}}\biggl[\ln\left(\frac{r^2-\sqrt{2q_m}\beta^{1/4}r+q_m\sqrt{\beta}}{r^2
+\sqrt{2q_m}\beta^{1/4}r+q_m\sqrt{\beta}}\right)
\]
\begin{equation}
+2\arctan\left(\frac{\sqrt{q_m}\beta^{1/4}+\sqrt{2}r}{\sqrt{q_m}\beta^{1/4}}\right)
-2\arctan\left(\frac{\sqrt{q_m}\beta^{1/4}-\sqrt{2}r}{\sqrt{q_m}\beta^{1/4}}\right)\biggr]-\frac{r^3}{2G_Nl^2}.
\label{10}
\end{equation}
The BH magnetic mass is given by \cite{Kr2}
\begin{equation}
m_M=4\pi\int_0^\infty \frac{q_m^2}{8\pi(r^4+\beta q_m^2)}r^2dr=\frac{\pi q_m^{3/2}}{4\sqrt{2}\beta^{1/4}}
\approx 0.56\frac{q_m^{3/2}}{\beta^{1/4}}.
\label{11}
\end{equation}
Equation (11) shows that at the limit $\beta=0$ (Maxwell's case) the magnetic energy becomes infinite.
Making use of Eqs. (7) and (10) we obtain the metric function
\begin{equation}
f(r)=1-\frac{2m_0G_N}{r}-\frac{q_m^{3/2}G_Ng(r)}{4\sqrt{2}\beta^{1/4}r}+\frac{r^2}{l^2},
\label{12}
\end{equation}
where
\[
g(r)\equiv \ln\left(\frac{r^2-\sqrt{2q_m}\beta^{1/4}r+q_m\sqrt{\beta}}{r^2+\sqrt{2q_m}\beta^{1/4}r+q_m\sqrt{\beta}}\right)
\]
\[
+2\arctan\left(\frac{\sqrt{q_m}\beta^{1/4}+\sqrt{2}r}{\sqrt{q_m}\beta^{1/4}}\right)
-2\arctan\left(\frac{\sqrt{q_m}\beta^{1/4}-\sqrt{2}r}{\sqrt{q_m}\beta^{1/4}}\right).
\]
In general relativity without the cosmological constant (neglecting $r^2/l^2$ in Eq. (12)) the metric function as $r\rightarrow \infty$ approaches to \cite{Kr2}
\begin{equation}
f(r)=1-\frac{2(m_0+m_M)G_N}{r}+\frac{q_m^2G_N}{r^2}+\mathcal{O}(r^{-5})~~~\mbox{as}~r\rightarrow \infty.
\label{13}
\end{equation}
In this case the correction to the Reissner--Nordstr\"{o}m solution, according to Eq. (13), is in the order of
$\mathcal{O}(r^{-5})$. The total mass (ADM mass) of the black hole $M\equiv m_0+m_M$ includes the Schwarzschild mass $m_0$ and the magnetic mass $m_M$.
It is worth noting that if we put $m_0=0$, i.e the black hole mass is the magnetic mass,
as $r\rightarrow 0$, from Eq. (12) we find the asymptotic with a de Sitter core
\begin{equation}
f(r)=1-\frac{G_Nr^2}{\beta}+\frac{r^2}{l^2}+\frac{G_Nr^6}{7\beta^2q_m^2}-\frac{G_Nr^{10}}{11\beta^3q_m^4}+\mathcal{O}(r^{12})~~~\mbox{as}~r\rightarrow 0.
\label{14}
\end{equation}
The solution (14) is regular because at $r= 0$ we have $f(0)= 1$.
The plot of the metric function (12) is depicted in Fig. 1 at $m_0=0$, $G_N=1$, $l=10$ for different parameters $q_m$ and $\beta$.
\begin{figure}[h]
\includegraphics  {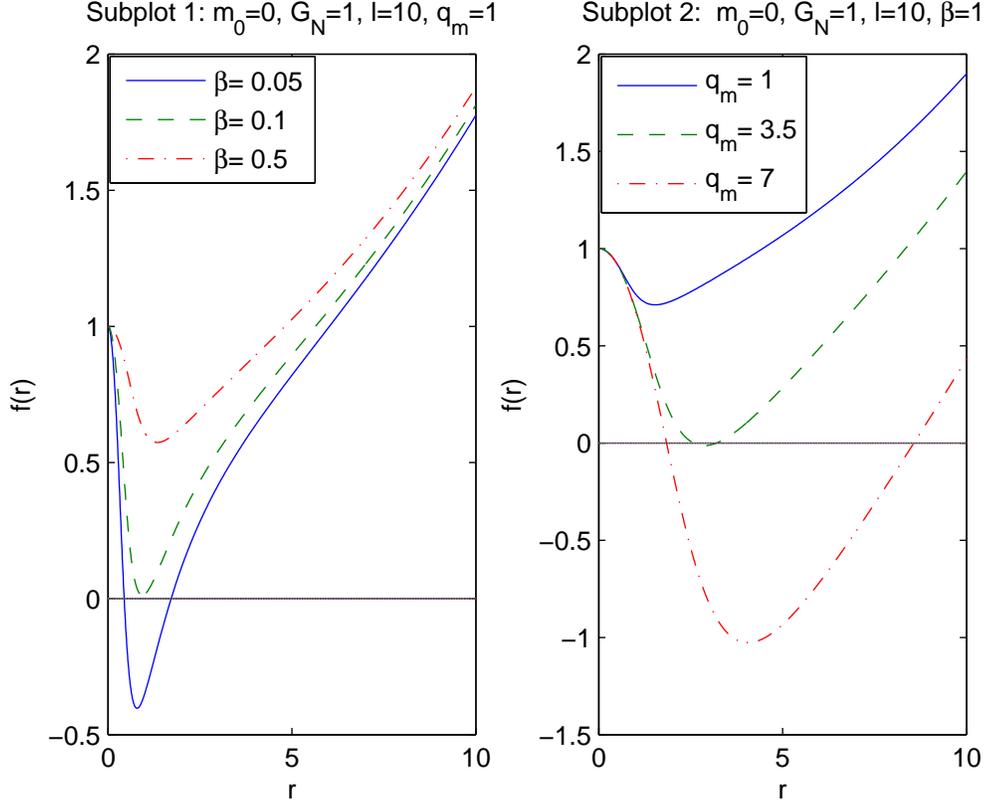}
\caption{\label{fig.1} The plot of the function $f(r)$ at $m_0=0$, $G_N=1$, $l=10$.}
\end{figure}
Figure 1  shows that black holes can have two horizons, one extreme horizon or no horizons depending on model parameters.

\section{First law of black hole thermodynamics and the Smarr relation}

If one considers that black holes are classical objects, then there is an analogy between black hole mechanics and thermodynamics. The role of the temperature plays the surface gravity and the event horizon corresponds to entropy $S$. In spite that classical black holes have zero temperature, Hawking proofed that black holes emit radiation with a blackbody spectrum \cite{Hawking}. The first law of black hole thermodynamics reads $\delta M=T\delta S+\Omega\delta J+\Phi\delta Q$ with black hole mass $M$, charge $Q$, and angular momentum $J$, where $M$, $J$, $Q$ are the extensive and $T$, $\Omega$, $\Phi$ are intensive thermodynamic variables. The disadvantage of formulated first law of black hole thermodynamics was the absence of the pressure-volume term $P\delta V$. To improve the first law of black hole thermodynamics the pressure was associated with a
negative cosmological constant $\Lambda$ which gives a positive vacuum pressure in space-time. Then the  generalized first law of black hole thermodynamics became $\delta M=T\delta S+ V\delta P+\Omega\delta J+\Phi\delta Q$ with $V=\partial M/\partial P$ at constant $S$, $J$, $Q$ \cite{Kastor}, \cite{Dolan1}, \cite{Cvetic}. Comparing the first law of black hole mechanics with ordinary thermodynamics requires to interpret $M$ as a chemical enthalpy \cite{Kastor}, $M=U+PV$, where $U$ is the internal energy.

To get the Smarr formula from the first law of BH thermodynamics we consider dimensions of physical quantities \cite{Smarr} (see also \cite{Kastor}). Let us consider the units with $G_N=1$. Then from the dimensional analysis we obtain $[M]=L$, $[S]=L^2$, $[P]=L^{-2}$, $[J]=L^2$, $[q_m]=L$, $[\beta]=L^2$. Considering $\beta$ as a thermodynamic variable and
taking into consideration the Euler’s theorem (see \cite{Mann}) (the Euler scaling argument), we obtain the mass $M(S,P,J,q_m,\beta)$,
\begin{equation}
M=2S\frac{\partial M}{\partial S}-2P\frac{\partial M}{\partial P}+2J\frac{\partial M}{\partial J}+q_m\frac{\partial M}{\partial q_m}+2\beta\frac{\partial M}{\partial \beta},
\label{15}
\end{equation}
where $\partial M/\partial \beta\equiv {\cal B}$ is the thermodynamic conjugate to the coupling $\beta$ and the black hole volume $V$ and pressure $P$ are given by \cite{Myers}, \cite{Myers1}
\begin{equation}
V=\frac{4}{3}\pi r_+^3,~~~P=-\frac{\Lambda}{8\pi}=\frac{3}{8\pi l^2}.
\label{16}
\end{equation}
We consider the non-rotating  stationary black hole so that $J=0$. From Eq. (12) and equation $f(r_+)=0$, where $r_+$ is the horizon radius, we find (at $G_N=1$)
\begin{equation}
M=\frac{r_+}{2}+\frac{r_+^3}{2l^2}+\frac{\pi q_m^{3/2}}{4\sqrt{2}\beta^{1/4}}-\frac{q_m^{3/2}g(r_+)}{8\sqrt{2}\beta^{1/4}}.
\label{17}
\end{equation}
Making use of Eq. (17) we obtain
\[
\delta M=\left(\frac{1}{2}+\frac{3r_+^2}{2l^2}-\frac{q_m^{3/2}}{8\sqrt{2}\beta^{1/4}}\frac{\delta g(r_+)}{\delta r_+}\right)\delta r_+
-\frac{r_+^3}{l^3}\delta l
\]
\[
+\left(\frac{q_m^{3/2}g(r_+)}{32\sqrt{2}\beta^{5/4}}-\frac{\pi q_m^{3/2}}{16\sqrt{2}\beta^{5/4}}-\frac{q_m^{3/2}}{8\sqrt{2}\beta^{1/4}}\frac{\delta g(r_+)}{\delta\beta}\right)\delta\beta
\]
\begin{equation}
+\left(\frac{3\pi q_m^{1/2}}{8\sqrt{2}\beta^{1/4}}-\frac{3q_m^{1/2}g(r_+)}{16\sqrt{2}\beta^{1/4}}-\frac{q_m^{3/2}}{8\sqrt{2}\beta^{1/4}}\frac{\delta g(r_+)}{\delta q_m}\right)\delta q_m.
\label{18}
\end{equation}
The Hawking temperature is given by
\begin{equation}
T=\frac{f'(r)|_{r=r_+}}{4\pi}.
\label{19}
\end{equation}
From Eqs. (12) and (19) we obtain the Hawking temperature ($G_N=1$)
\begin{equation}
T=\frac{1}{4\pi}\biggl(\frac{1}{r_+}+\frac{3r_+}{l^2}-\frac{q_m^2r_+}{r_+^4+\beta q_m^2}\biggr),
\label{20}
\end{equation}
where we have used the equation
\begin{equation}
\frac{\delta g(r_+)}{\delta r_+}=\frac{4\sqrt{2q_m}\beta^{1/4}r_+^2}{r^4+q_m^2\beta}.
\label{21}
\end{equation}
Making use of Eqs. (17), (20) and (21) we obtain
\begin{equation}
\frac{\partial M(r_+)}{\partial r_+}=2\pi r_+T.
\label{22}
\end{equation}
Then we find the entropy in our case of RNED-AdS black hole
\begin{equation}
S=\int \frac{d M(r_+)}{T}=\int \frac{1}{T}\frac{\partial M(r_+)}{\partial r_+}dr_+=\pi r_+^2.
\label{23}
\end{equation}
Thus, the Bekenstein--Hawking entropy holds. Then from relations
\[
\frac{\delta g(r_+)}{\delta\beta}=-\frac{\sqrt{2q_m}r_+^3}{\beta^{3/4}(r_+^4+\beta q_m^2)} ,~~~\frac{\delta g(r_+)}{\delta q_m}=
-\frac{2\sqrt{2}\beta^{1/4}r_+^3}{q_m^{1/2}(r_+^4+\beta q_m^2)}
\]
and Eqs. (16), (18), (20)  and (23) we obtain the first law of black hole thermodynamics
\begin{equation}
\delta M = T\delta S + V\delta P + \Phi_m \delta q_m + {\cal B}\delta\beta,
\label{24}
\end{equation}
where the thermodynamic \textit{magnetic} potential $\Phi_m$ and the thermodynamic conjugate to the coupling $\beta$ are given by
\[
\Phi_m =\frac{q_mr_+^3}{4(r_+^4+\beta q_m^2)}+\frac{3\pi q_m^{1/2}}{8\sqrt{2}\beta^{1/4}}- \frac{3q_m^{1/2}g(r_+)}{16\sqrt{2}\beta^{1/4}},
\]
\begin{equation} {\cal B}=\frac{q_m^{3/2}g(r_+)}{32\sqrt{2}\beta^{5/4}}-\frac{\pi q_m^{3/2}}{16\sqrt{2}\beta^{5/4}}+
 \frac{q_m^2r_+^3}{8\beta(r_+^4+\beta q_m^2)}.
\label{25}
\end{equation}
The ${\cal B}$ in the Born--Infeld AdS case was referred as `Born--Infeld vacuum polarization' \cite{Mann1}. The presence of ${\cal B}$ is needed for a consistency of the Smarr formula. The plot of $\Phi_m$ vs. $r_+$ is depicted in Fig. 2.
\begin{figure}[h]
\includegraphics {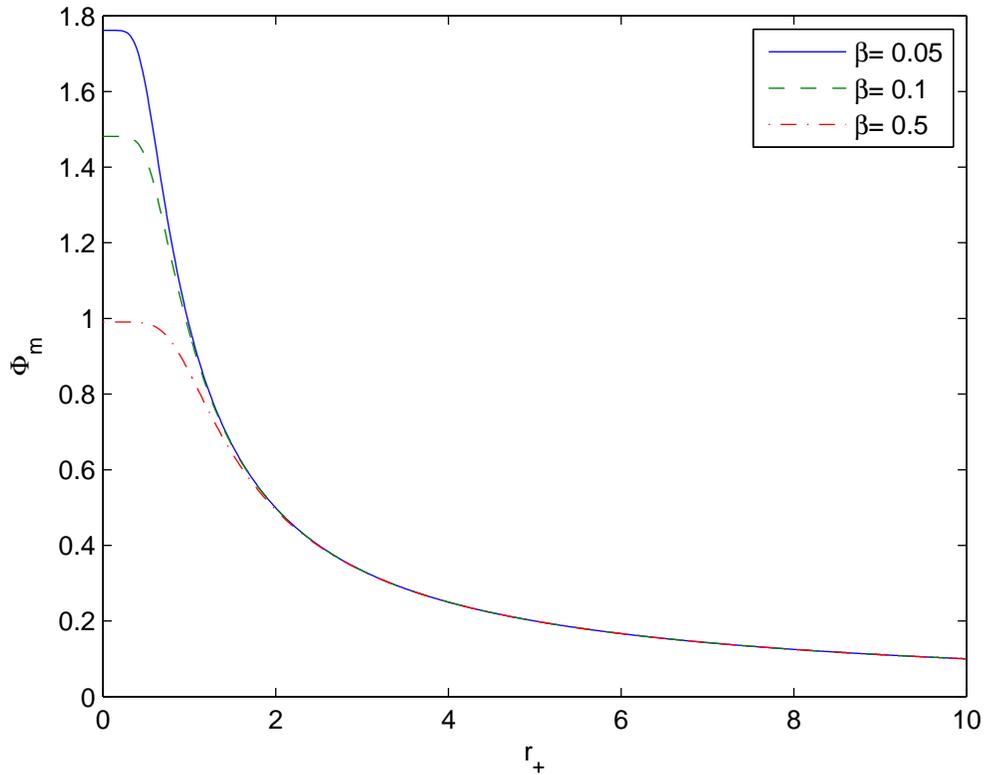}
\caption{\label{fig.2} The plot of the function $\Phi_m$ vs. $r_+$ at $q_m=1$. The solid curve is for $\beta=0.05$, the dashed curve corresponds to $\beta=0.1$, and the dashed-doted curve corresponds to $\beta=0.5$.}
\end{figure}
 According to Fig. 2 when coupling $\beta$ increases the magnetic potential decreases and at $r_+\rightarrow \infty$ it vanishes, $\Phi_m(\infty)=0$. At $r_+ = 0$ $\Phi_m$ is the finite value.
\begin{figure}[h]
\includegraphics{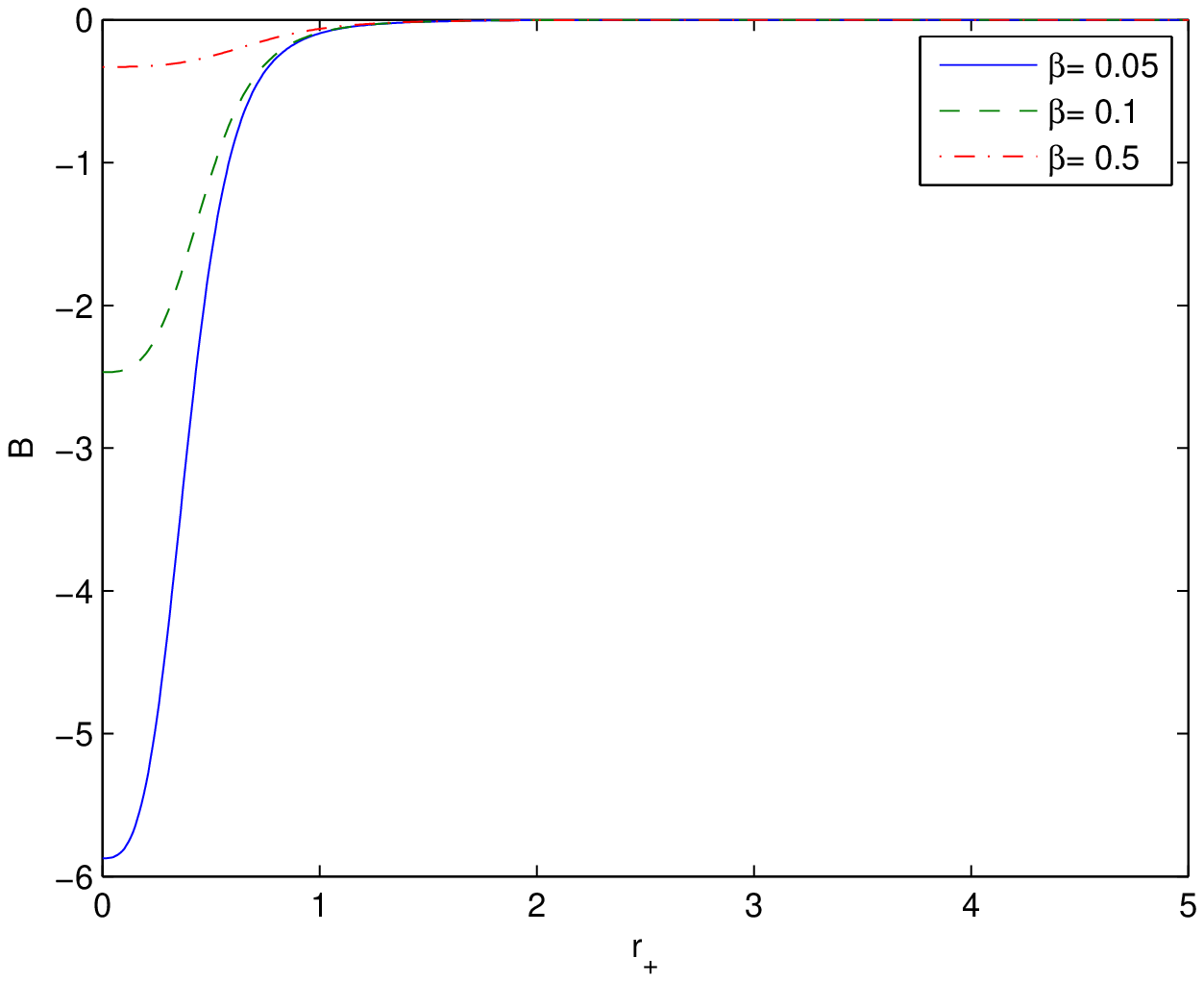}
\caption{\label{fig.3} The plot of vacuum polarization ${\cal B}$ vs. $r_+$ at $q_m=1$. The solid curve is for $\beta=0.05$, the dashed curve corresponds to $\beta=0.1$, and the dashed-doted curve corresponds to $\beta=0.5$.}
\end{figure}
The plot of the function ${\cal B}$ vs. $r_+$ is presented in Fig. 3. It is worth noting that at $r_+ = 0$ the vacuum polarization ${\cal B}$ is finite. Figure 3 shows that when coupling $\beta$ increases the absolute value of vacuum polarization decreases and at $r_+\rightarrow \infty$ it becomes zero, ${\cal B}(\infty)=0$.

Making use of Eqs. (16), (17), (23) and (25), one can verify that the generalized Smarr formula holds,
\begin{equation}
M=2ST-2PV+q_m\Phi_m+2\beta{\cal B}.
\label{26}
\end{equation}
Making use of the Bekenstein and Hawking arguments \cite{Bekenstein}, \cite{Hawking1}, we conclude that the second law of thermodynamics for AdS black holes also holds. The study of Born--Infeld electrodynamics in AdS space-time in the extended phase space was presented in \cite{Zou}, \cite{Hendi}, \cite{Hendi1}, \cite{Zeng}.

\section{The black hole thermodynamics}

Making use of Eq. (20) we obtain the equation of state (EoS) for RNED AdS black hole
\begin{equation}
P=\frac{T}{2r_+}-\frac{1}{8\pi r_+^2}+\frac{q_m^2}{8\pi(r_+^4+\beta q_m^2)}.
\label{27}
\end{equation}
At $\beta=0$ Eq. (27) becomes EoS for a charged (by linear Maxwell electrodynamics) AdS black hole \cite{Mann2}.
If one compare EoS of charged AdS black hole with the Van der Waals equation, then the specific volume $v$ should be identified
with $2l_Pr_+$ \cite{Mann2}. With $l_P=\sqrt{G_N}=1$ the horizon diameter $2r_+$ plays the role of the specific volume of the corresponding fluid.  Thus, Eq. (27) becomes
\begin{equation}
P=\frac{T}{v}-\frac{1}{2\pi v^2}+\frac{2q_m^2}{\pi(v^4+16\beta q_m^2)}.
\label{28}
\end{equation}
 Equation (28) qualitatively mimics the behaviour of the Van der Waals fluid. Critical points take place at the inflection in the $P-v$ diagram with
\[
\frac{\partial P}{\partial v}=-\frac{T}{v^2}+\frac{1}{\pi v^3}-\frac{8q_m^2v^3}{\pi(v^4+16\beta q_m^2)^2}=0,
\]
\begin{equation}
\frac{\partial^2 P}{\partial v^2}=\frac{2T}{v^3}-\frac{3}{\pi v^4}-\frac{8q_m^2v^2(48\beta q_m^2-5v^4)}{\pi(v^4+16\beta q_m^2)^3}=0.
\label{29}
\end{equation}
From Eq. (29) we obtain the equation for the critical points as follows
\begin{equation}
8q_m^2v_c^6(3v_c^4-80\beta q_m^2)-(v_c^4+16\beta q_m^2)^3=0.
\label{30}
\end{equation}
It is difficult to obtain an analytic solution to Eq. (30). Equation (30) can be represented as the cubic equation for the parameter $\beta$ with the solution
\begin{equation}
\beta=\frac{\sqrt{5}v_c^3}{2\sqrt{6}q_m}\sinh\left(\frac{1}{3}\sinh^{-1}\left(\frac{3\sqrt{6}}{5\sqrt{5}q_m}v_c\right)\right)-\frac{v_c^4}{16q_m^2},
\label{31}
\end{equation}
where $\sinh^{-1}(x)$ is the inverse hyperbolic $\sinh$-function. Making use of Eq. (31), the function $v_c$ vs. $\beta $ with $q_m=1$ is depicted in Fig. 4.
\begin{figure}[h]
\includegraphics {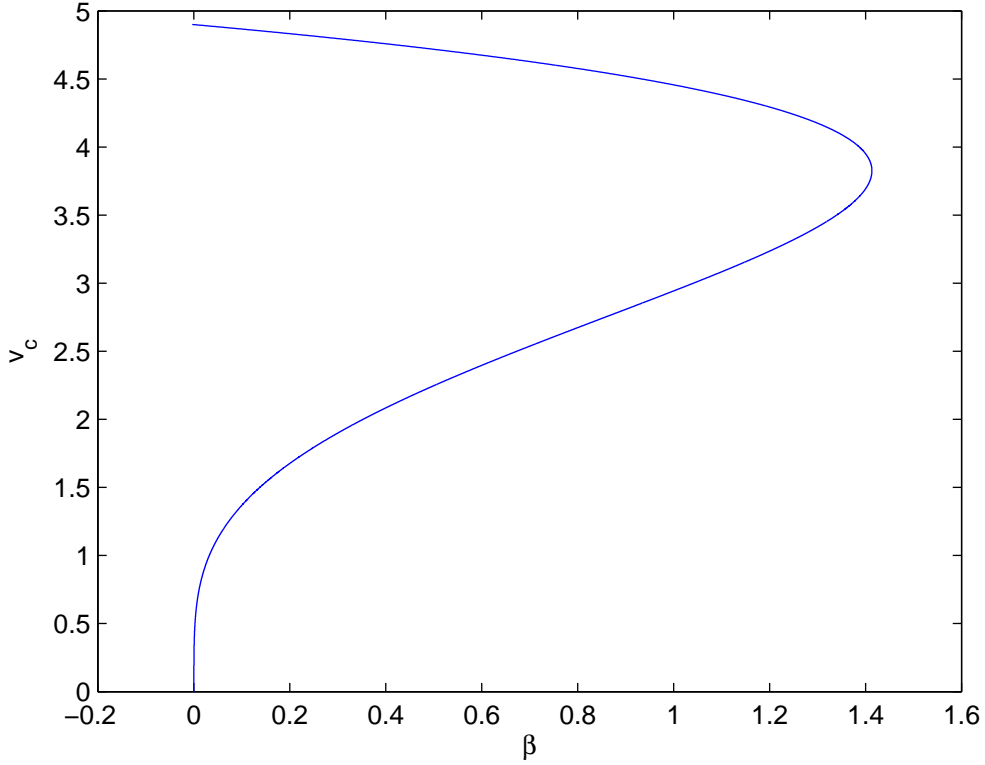}
\caption{\label{fig.4} The plot of the function $v_c$ vs. $\beta$ at $q_m=1$.}
\end{figure}
In accordance with Fig. 4 at $\beta >1.41$ (approximately), there are not real solutions to Eq. (30). At $0<\beta<1.41$, for each $\beta$ there are two real solutions for $v_c$. The expression for the critical temperature follows from Eq. (27)
\begin{equation}
T_c=\frac{1}{\pi v_c}-\frac{8q_m^2v_c^5}{\pi(v_c^4+16\beta q_m^2)^2}.
\label{32}
\end{equation}
Numerical solutions to Eq. (30) as well as the critical temperatures for different values of $\beta$ are presented in Table 1 and showing two inflection points for each $\beta$.
\begin{table}[ht]
\caption{Critical values of the specific volume and temperature at $q_m=1$}
\centering
\begin{tabular}{c c c c c c c c c c}\\[1ex]
\hline
$\beta $ & 0.6 & 0.7 & 0.8 & 0.9 & 1 & 1.1 & 1.2 & 1.3 & 1.4  \\[0.5ex]
\hline
$v_{c1}$ &2.397 & 2.538 & 2.674 & 2.808 & 2.943 & 3.084 & 3.235 & 3.413 & 3.693 \\[0.5ex]
\hline
$T_{c1}$ &0.0218 & 0.0288 & 0.0339 & 0.0375 & 0.0403 & 0.0424 & 0.0439 & 0.0451 & 0.04592 \\[0.5ex]
\hline
$v_{c2}$ & 4.675 & 4.628 & 4.577 & 4.4557 & 4.457 & 4.383 & 4.294 & 4.175 & 3.951 \\[0.5ex]
\hline
$T_{c2}$ &0.0441 & 0.0443 & 0.0445 & 0.0447 & 0.0449 & 0.0451 & 0.0453 & 0.0456 & 0.04594 \\[0.5ex]
\hline
\end{tabular}
\end{table}
The plot of $T_c$ vs. $\beta$ is depicted in Fig. 5.
\begin{figure}[h]
\includegraphics{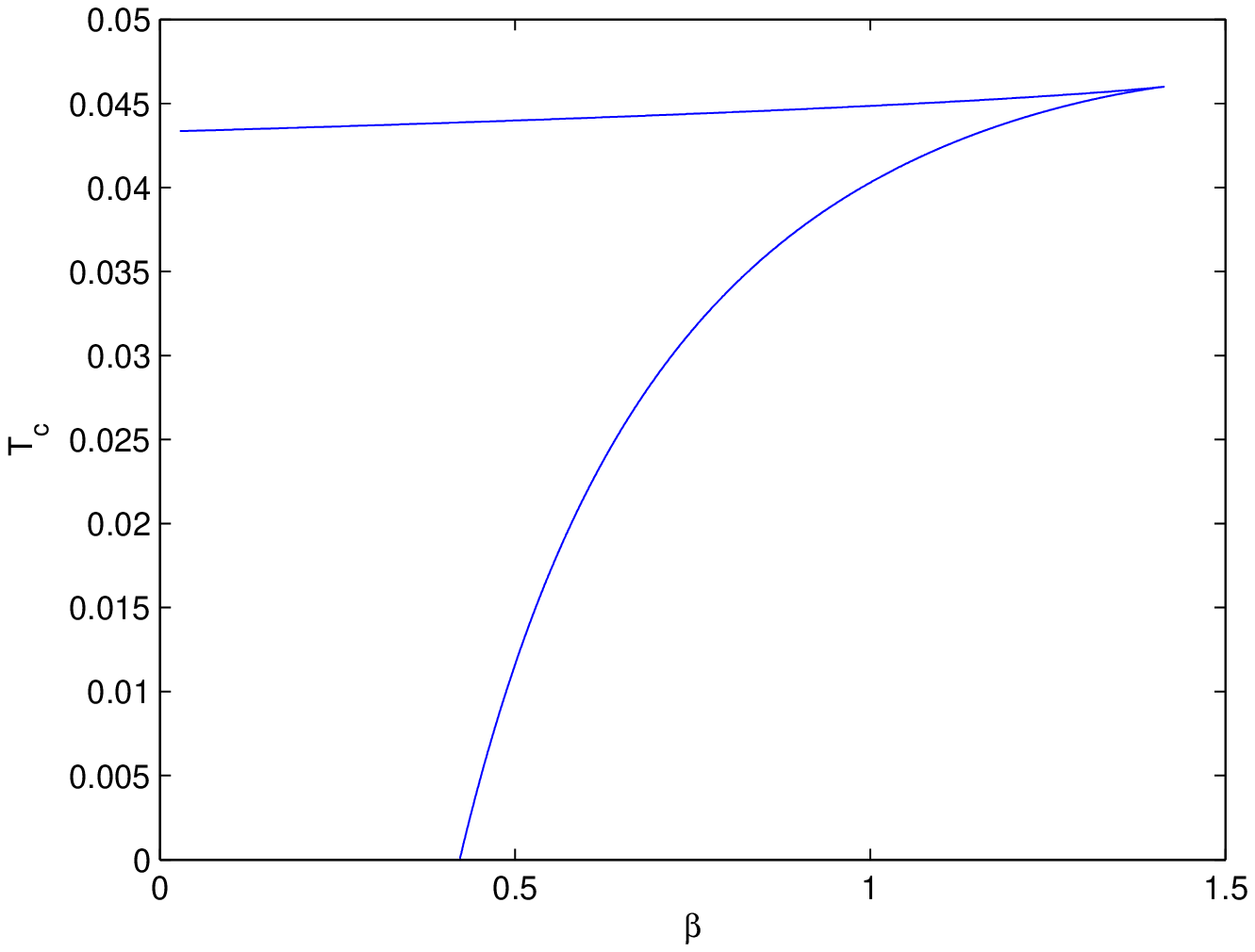}
\caption{\label{fig.5} The plot of the critical temperature $T_c$ vs. $\beta$ at $q_m=1$.}
\end{figure}
In accordance with Fig. 5, at $0<\beta<0.42$ (approximately) there is one critical temperature (for each $\beta$) but for $0.42<\beta<1.41$ two.
It is worth noting that also Fig. 4 shows that for each $\beta$ there are two real positive critical points $v_c$ but for the interval $0<\beta<0.42$ only one $v_c$  gives the physical positive critical temperature.
At the point $v_c$ we have second-order phase transition.
The $P-v$ diagrams are given in Figs. 6 and 7 for some values of $T$.
\begin{figure}[h]
\includegraphics {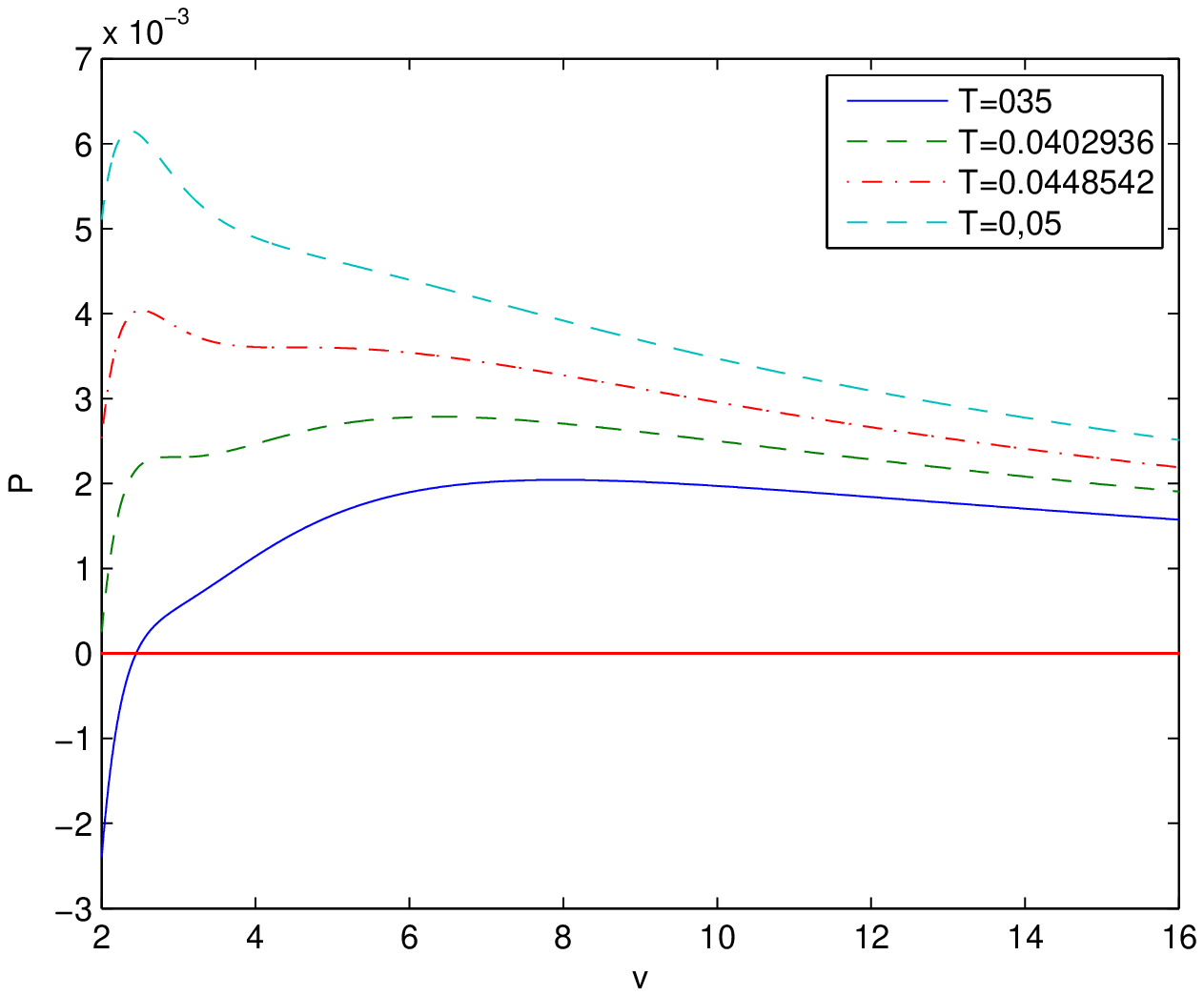}
\caption{\label{fig.6} The plot of the function $P$ vs. $v$ at $q_m=\beta=1$. The critical isotherms correspond to $T_{c1}=0.0402936$ and
$T_{c2}=0.0448542$.}
\end{figure}
\begin{figure}[h]
\includegraphics {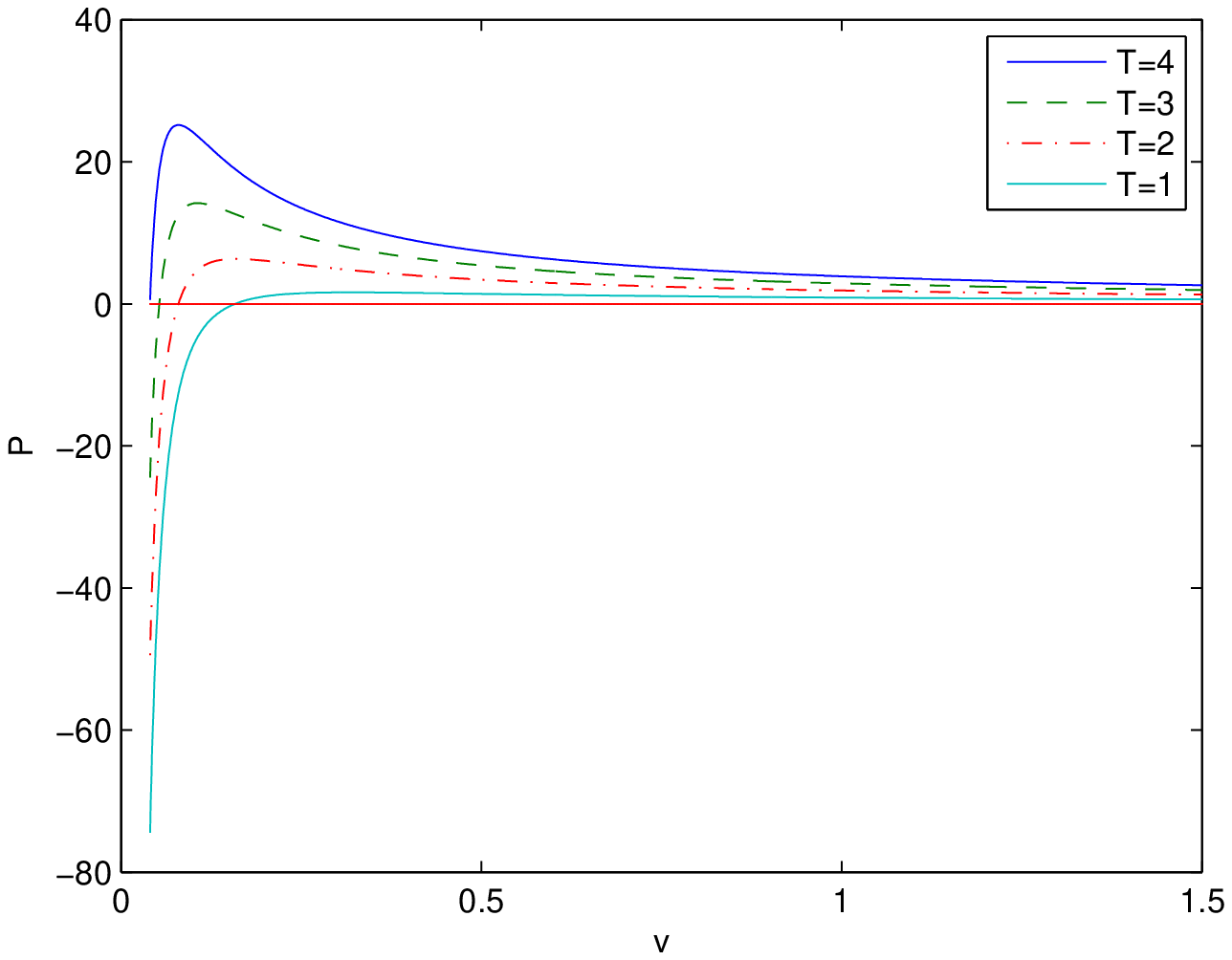}
\caption{\label{fig.7} The plot of the function $P$ vs. $v$ at $q_m=\beta=1$ for $T=1, 2, 3, 4$.}
\end{figure}
According to Fig. 6 for $q_m=\beta=1$ there are two critical values $v_{c1}\approx 2.94305$ ($T_{c1}=0.0402936$) and $v_{c2}\approx 4.45663$ ($T_{c2}=0.0448542$). Thus, there are inflection points and the EoS in our case is more complicated compared to the Van der Waals gas EoS and similar to Born--Infeld AdS case. Figure 7 shows non-critical behaviour of $P-v$ diagrams for $T=1, 2, 3$ and $4$.
The critical pressure is given by
\begin{equation}
P_c=\frac{1}{2\pi v_c^2}+\frac{2q_m^2(16\beta q_m^2-3v_c^4)}{\pi(v_c^4+16\beta q_m^2)^2}.
\label{33}
\end{equation}
 The plot of $P_c$ vs. $\beta$ is presented in Fig. 8.
 \begin{figure}[h]
\includegraphics {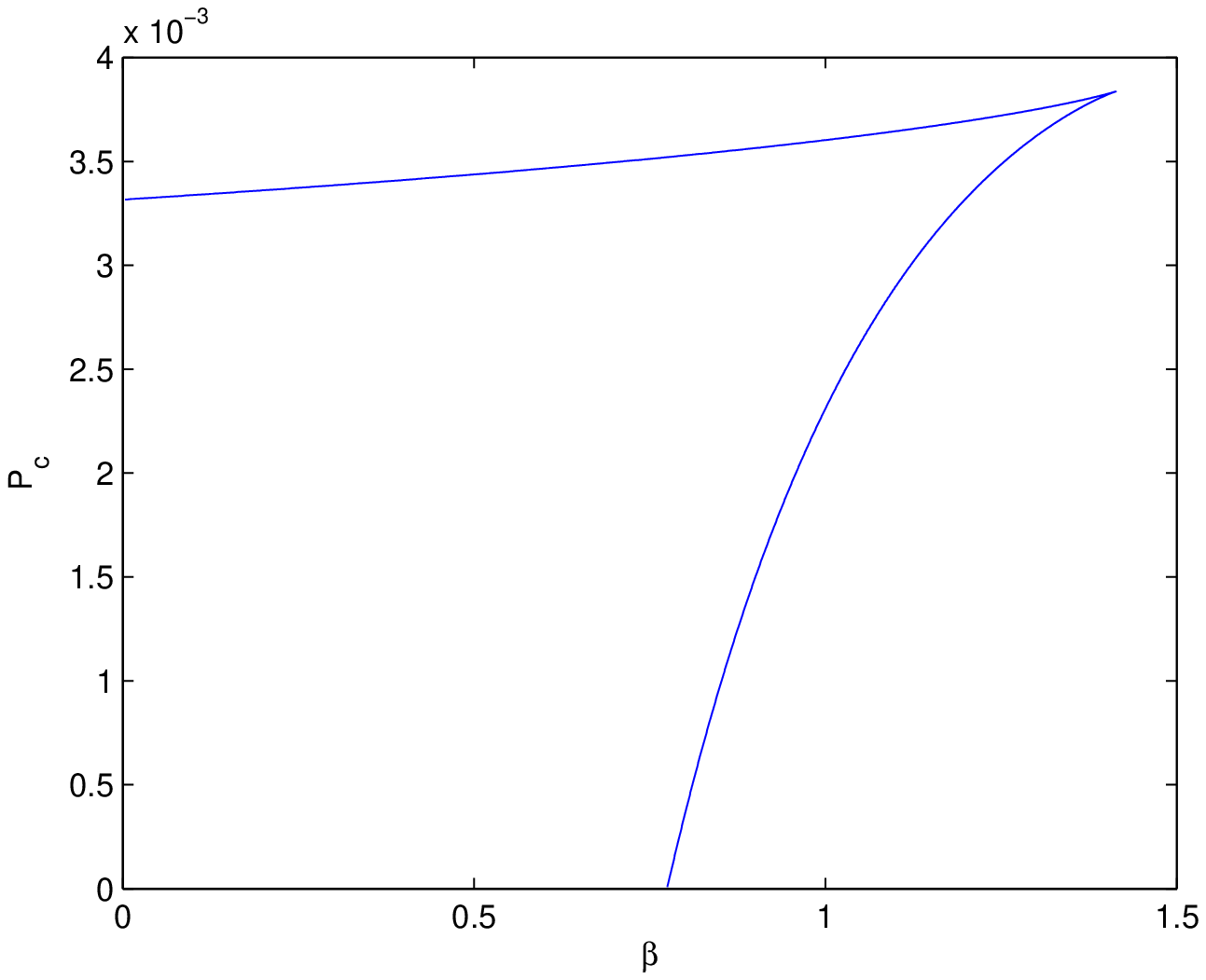}
\caption{\label{fig.8} The plot of the critical pressure $P_c$ vs. $\beta$ at $q_m=1$.}
\end{figure}
According to Fig. 8, at $0<\beta<0.77$ (approximately) there is one critical pressure (for each $\beta$) but for $0.77<\beta<1.41$ two.
Also for each $\beta$ there are two real positive solutions to Eq. (30) for critical points $v_c$, but for the interval $0<\beta<0.77$ only one $v_c$  gives the physical positive critical pressure. Making use of Eqs. (32) and (33) one obtains the critical ratio
\begin{equation}
\rho_c=\frac{P_cv_c}{T_c}=\frac{(v_c^4+16\beta q_m^2)^2+4q_m^2v_c^2(16\beta q_m^2-3v_c^4)}{2[(v_c^4+16\beta q_m^2)^2-8q_m^2v_c^6]},
\label{34}
\end{equation}
where $\beta$ is given by Eq. (31). The plot of $\rho_c$ vs. $\beta$ at $q_m=1$ is depicted in Fig. 9.
\begin{figure}[h]
\includegraphics {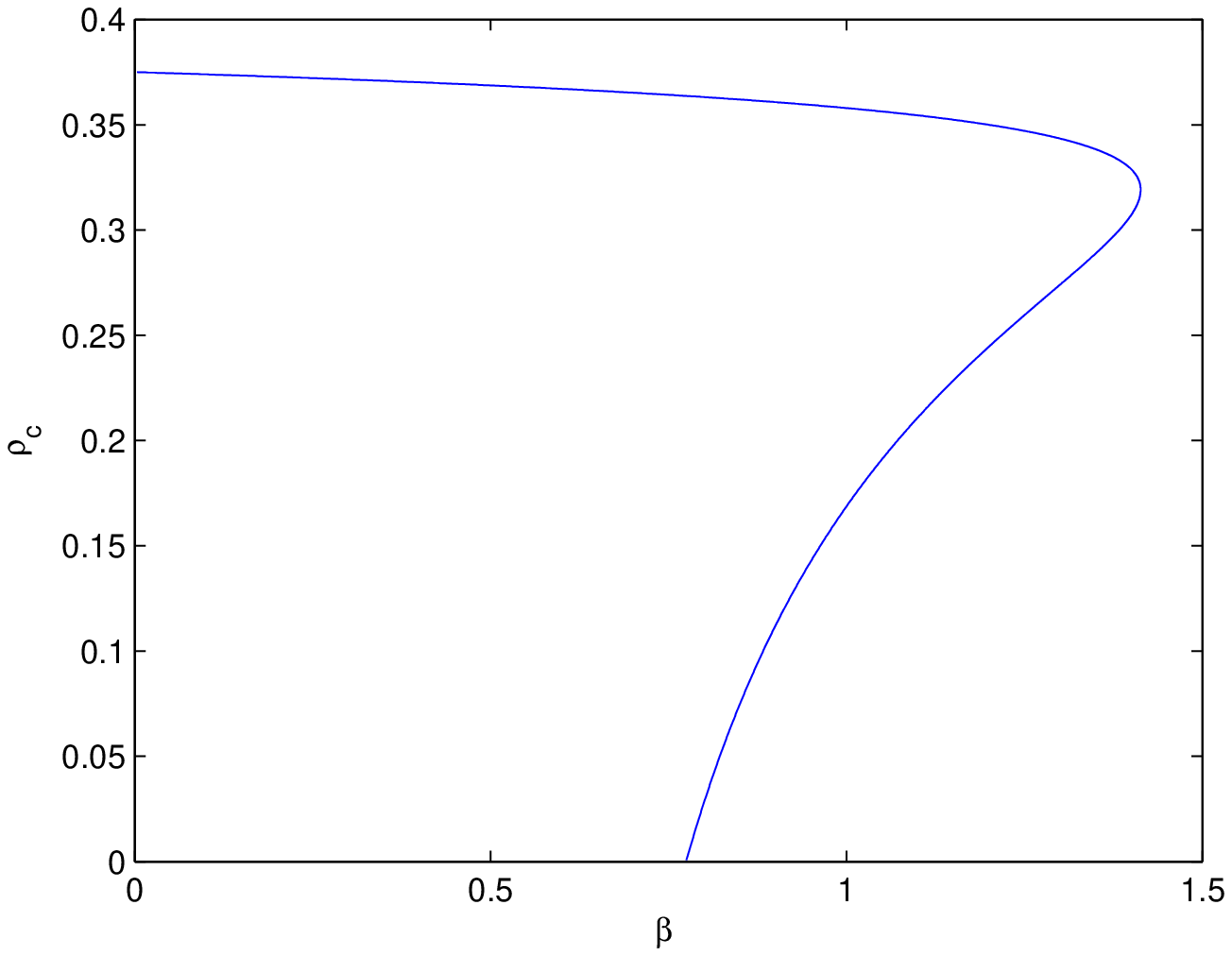}
\caption{\label{fig.9} The plot of the critical ratio $\rho_c$ vs. $\beta$.}
\end{figure}
At $\beta=0$ we get $\rho_c=3/8$ as for a Van der Waals fluid. In accordance with Fig. 9 the critical ratio
in our model decreases with $\beta$.

\subsection{ The Gibbs free energy}

Let us consider the expression for the Gibbs free energy for a fixed charge, coupling $\beta$ and pressure
\begin{equation}
G=M-TS.
\label{35}
\end{equation}
Here, $M$ is considered as a chemical enthalpy, that is the total energy of a system with its internal energy $U$ and the energy $PV$ to displace the vacuum energy of its environment: $M =U+PV$. From Eqs. (16) and (35) (at $G_N=1$) we obtain
\begin{equation}
G=\frac{r_+}{2}+\frac{4\pi r_+^3P}{3}+\frac{\pi q_m^{3/2}}{4\sqrt{2}\beta^{1/4}}-\frac{q_m^{3/2}g(r_+)}{8\sqrt{2}\beta^{1/4}}
-\pi Tr_+^2,
\label{36}
\end{equation}
where $r_+$ is a function of $P$ and $T$ (see Eq. (27)). The plot of $G$ vs. $T$ is depicted in Fig. 10.
\begin{figure}[h]
\includegraphics {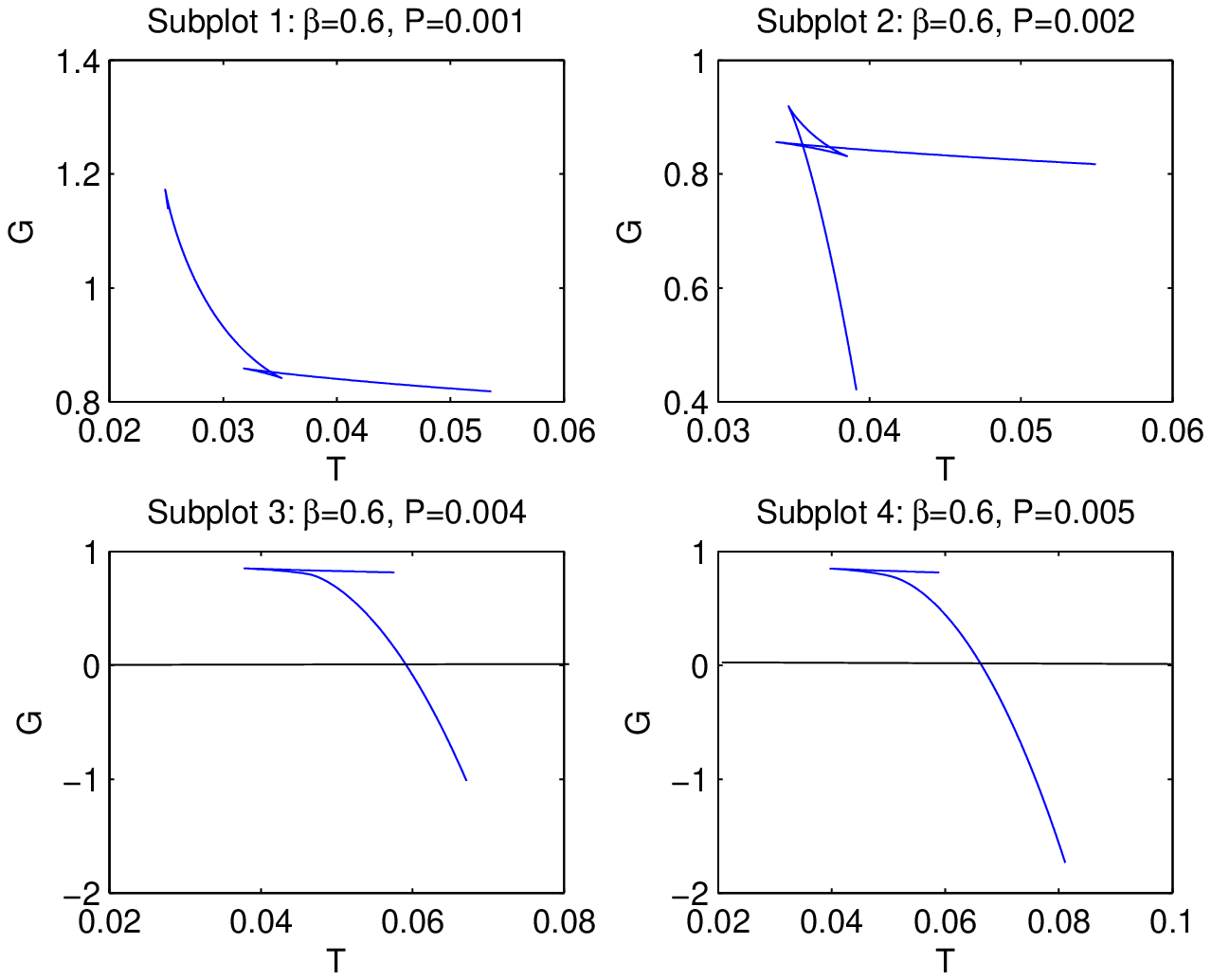}
\caption{\label{fig.10} The plot of the Gibbs free energy $G$ vs. $T$ with $q_m=1$, $\beta=0.6$.}
\end{figure}
The behaviour of $G$ depends on pressure $P$ and coupling $\beta$. As an example we consider the case with $\beta=0.6$ where there is one physical critical point (see Table 1 and Figs. 5 and 8) $v_c\approx 4.6745$, $T_c\approx 0.0441$, $P_c\approx 0.0035$. The behaviour of the Gibbs free energy is similar to the RN-AdS black hole with one critical point and the corresponding first-order phase transition between small and large black holes (in subplots 1 and 2). In this case there is a point at which two black holes have equal free energy. One can see two branches of black holes with a cusp and the Gibbs free energy shows 'swallowtail' behaviour with a first-order phase transition between two branches for $P<P_c$. Subplots 3 and 4 in Fig. 10 displace a characteristic shape similar to the Hawking--Page behavior for Schwarzschild-AdS case and there is no first-order phase transition in the system for $P>P_c$.

\subsection{Critical exponents}

We expand the critical values in small parameter $\beta$
\[
v_c=2\sqrt{6}q_m-\frac{7}{9\sqrt{6}}\frac{\beta}{q_m}+{\cal O}(\beta^2),~~~
T_c=\frac{1}{3\sqrt{6}\pi q_m}+\frac{1}{108\sqrt{6}\pi}\frac{\beta}{q_m^3}+{\cal O}(\beta^2),~~~
\]
\begin{equation}
P_c=\frac{1}{96\pi q_m^2}+\frac{7}{10368\pi}\frac{\beta}{q_m^4}+{\cal O}(\beta^2).
\label{37}
\end{equation}
It is worth noting that the critical point (37) at $\beta=0$ is the same as in charged AdS black hole \cite{Mann1}, but there are corrections due to coupling $\beta$. The critical ratio $\rho_c$ vs. $\beta$ is depicted in Fig. 9 and the analytical expression for small $\beta$ is given by
\begin{equation}
\rho_c=\frac{3}{8}+\frac{1}{96}\frac{\beta}{q_m^2}+{\cal O}(\beta^2).
\label{38}
\end{equation}
The value $\rho_c=3/8$ takes place for the Van der Waals fluid. The critical exponents show the physical quantity behaviour in the vicinity of the critical points which do not depend on details of the system. The exponent $\alpha$ defines the behaviour of the specific heat at the constant volume
\begin{equation}
C_v=T\frac{\partial S}{\partial T}\propto |t|^{-\alpha},
\label{39}
\end{equation}
where $t=(T-T_c)/T_c$. Because the entropy $S=\pi r_+^2=(3V/(4\pi))^{2/3}$ is constant, we have $C_v=0$ and, therefore, $\alpha=0$.
Let us define the quantities \cite{Mann}
\begin{equation}
p=\frac{P}{P_c},~~\nu=\frac{v}{v_c}=\sqrt[3]{\omega+1},~~\tau=\frac{T}{T_c}=t+1.
\label{40}
\end{equation}
Taking into account Eq. (28), we obtain
\begin{equation}
p=\frac{\tau}{\nu \rho_c}-\frac{1}{2\pi \nu^2P_cv_c^2}+\frac{2q_m^2}{\pi P_c(\nu^4v_c^4+16\beta q_m^2)},
\label{41}
\end{equation}
where $P_c$ is given by Eq. (33).
One can expand $p$ in small parameters $t$ and $\omega$ near the critical point
\begin{equation}
p=1+At-Bt\omega -C\omega^3-Dt\omega^2+{\cal O}(\omega^4),
\label{42}
\end{equation}
where
\[
A=\frac{1}{\rho_c},~~~~B=\frac{1}{3\rho_c},~~~~D=-\frac{2}{9\rho_c},
\]
\begin{equation}
C=\frac{14}{81 \rho_c}-\frac{20}{81\pi P_cv_c^2}-\frac{8q_m^2v_c^8[(16\beta q_m^2)^2+224\beta q_m^2v_c^4-35v_c^8]}{81\pi P_cv_c^4(v_c^4+16\beta q_m^2)^4}.
\label{43}
\end{equation}
We included in Eq. (42) the additional term $Dt\omega^2$, compared to \cite{Mann}, which is in the same order as $\omega^3$.
The small $\beta$ expansion gives
\[
C=\frac{4}{81}+{\cal O}(\beta).
\]
It is worth noting that value 4/81 is realized in the RN-AdS case \cite{Mann}.
We will follow the same avenue as in \cite{Mann} to obtain critical exponents. Making use of Eq. (40) we obtain
\begin{equation}
dP=-P_c(Bt+2Dt\omega+3C\omega^2)d\omega.
\label{44}
\end{equation}
By using Maxwell’s equal area law \cite{Mann1}, one finds
\cite{Mann}
\begin{equation}
\omega_l(Bt+Dt\omega_l+C\omega_l^2)=\omega_s(Bt+Dt\omega_s+C\omega_s^2),
\label{45}
\end{equation}
\begin{equation}
\int_{\omega_l}^{\omega_s} \omega dP=0,
\label{46}
\end{equation}
where $\omega_s$ and $\omega_l$ correspond to the small and large black holes. The solution to Eqs. (45) and (46) is given by
\begin{equation}
\omega_l=\frac{-Dt+\sqrt{D^2t^2-4BCt}}{2C},~~~\omega_s=\frac{-Dt-\sqrt{D^2t^2-4BCt}}{2C}.
\label{47}
\end{equation}
At $D=0$ Eq. (47) becomes the solution obtained in \cite{Mann1}. Equation (47) is satisfied in the leading order up to ${\cal O} (t^{5/2})$.
We use the definitions: the difference of the large and small black hole volume on the given isotherm $v_l-v_s$, isothermal compressibility $\kappa_T$, $|P-P_c|$ on the critical isotherm $T = T_c$,
\begin{equation}
\eta=v_l-v_s\propto |t|^\beta,~~\kappa_T=-\frac{1}{v}\frac{\partial v}{\partial P}|_T\propto |t|^{-\gamma},~~|P-P_c|\propto |v-v_c|^\delta.
\label{48}
\end{equation}
Following the procedure of \cite{Mann1}, one obtains the same values of critical exponents as in the BI-AdS case
\begin{equation}
\beta=\frac{1}{2},~~~~\gamma=3,~~~~\delta=3.
\label{49}
\end{equation}
We have studied critical exponents in the vicinity of the critical point for a small non-linearity parameter $\beta$ and obtained the result as in the mean field theory. Thus, we have the same universality class as for the Van der Waals fluid. When parameter $\beta$ is not small, we can not expand critical temperature and pressure in $\beta$. Therefore, equalities in Eq. (40) will not hold and the non-linearity of electromagnetism will influence on the critical exponents.

\section{Summary}

We have studied the thermodynamic behaviour of charged by RNED black holes in an extended thermodynamic phase space. In this approach the cosmological constant is identified with a thermodynamic pressure and the mass of the black hole is the chemical enthalpy. We show the analogy with the Van der Walls liquid–gas system with the specific volume in the Van der Waals equation being the diameter of the event horizon (at $G_N=1$). The critical ratio $\rho_c = P_cv_c/T_c$ is equal to Van der Waals value of 3/8 plus corrections ${\cal O}(\beta)$ due to coupling $\beta$. The critical exponents coincide with those of the Van der Waals system similar to the BI-AdS case. The thermodynamics of RNED-AdS model was investigated showing the critical behaviour and phase transitions. The phase space includes the conjugate pair $({\cal B},\beta$). A thermodynamic quantity ${\cal B}$ conjugated to the nonlinear parameter $\beta$ of RNED has defined. We have demonstrated the consistency of the first law of black hole thermodynamics and the Smarr formula which depends on the quantities ${\cal B}$, $\Phi_m$ introduced. The critical points and phase transitions also depend on the RNED parameter $\beta$. Therefore, black hole thermodynamics (and black
hole physics) is modified in our model of RNED-AdS. The critical exponents were calculated which are the same as in the BI-AdS case.

\end{document}